\documentclass[12pt]{iopart}

\usepackage{setstack,iopams,xspace,graphicx,latexsym}

\newcommand{\order}[1]{\Or(#1)}

\newcommand{\ra}{\rightarrow}
\newcommand{\p}{\partial}
\newcommand{\om}{\omega}
\newcommand{\eps}{\epsilon}
\renewcommand{\vector}[1]{{\bf #1}}

\newcommand{\Z}{\ensuremath{\mathbb{Z}}\xspace}

\begin{document}

\title{Cylindrical Korteweg-de Vries solitons on a ferrofluid surface}

\author{Dirk Rannacher, Andreas Engel}
\address{Universit\"at Oldenburg, Institut f\"ur Theoretische
Physik,
26111 Oldenburg, Germany}
\ead{rannacher@theorie.physik.uni-oldenburg.de}

\begin{abstract}
Linear and non-linear surface waves on a ferrofluid cylinder
surrounding a current-carrying wire are investigated. Suppressing the 
Rayleigh-Plateau instability of the fluid column by the magnetic field
of a sufficiently large current in the wire axis-symmetric surface
deformations are shown to propagate without dispersion in the long
wavelength limit. Using multiple scale perturbation theory the weakly
non-linear regime may be described by a Korteweg-de~Vries equation
with coefficients depending on the magnetic field strength. For 
different values for the current in the wire hence different solutions
such as hump or hole solitons may be generated. The possibility to
observe these structures in experiments is also elucidated. 
\end{abstract}


\submitto{\NJP}

\maketitle


\section{Introduction}
Solitons are among the most interesting structures in
nature. Being configurations of continuous fields they retain
their localized shape even after interactions and collisions. Observed
originally long ago as stable moving humps in shallow water channels
\cite{ScRu} they have been established since then in 
various physical systems including optical waveguides, crystal
lattices, Josephson junctions, plasmas and spiral galaxies (for an
introduction see \cite{DrJo}). Long lasting efforts to theoretically
describe their intriguing properties have culminated in the
development of the inverse scattering technique \cite{GGKM} which is
among the most powerful methods to obtain exact solutions of nonlinear
partial differential equations \cite{AbSe}. 

Particularly popular examples for solitons in hydrodynamic systems
are the solutions of the Korteweg-de~Vries equation (KdV) 
\begin{equation}
  \label{eq:defkdv}
  \p_t u(x,t)+ 6u(x,t)\p_x u(x,t) + \p_x^3 u(x,t)=0 \, ,
\end{equation}
with $x$ standing for a space coordinate and $t$ denoting time. With $u$
representing the surface elevation of a liquid in a shallow duct this
equation can be derived perturbatively from the Euler equation
for the motion of an incompressible and inviscid fluid
\cite{KdV,whitham}. The one-soliton solution of (\ref{eq:defkdv}) is
given by  
\begin{equation}
  \label{eq:onesol}
  u(x,t)=\frac{c}{2}\, \mathrm{sech}^2\left(\frac{\sqrt{c}}{2}(x-ct)\right)
\end{equation}
which for all values of $c>0$ describes a hump of invariable shape
moving to the right with velocity $c$. The amplitude of the hump is
given by $c/2$ whereas $L=2/\sqrt{c}$ is a measure of its width. 

A decisive prerequisite to derive (\ref{eq:defkdv}) is that to linear
order in the field $u$ the system under consideration admits  
travelling waves $u\sim e^{i(kx-\om t)}$ with dispersion relation
\begin{equation}
  \label{eq:disp0}
  \om=c_0 k +\order{k^3} \qquad {\rm for} \qquad k\to 0\, ,
\end{equation}
where $c_0$ denotes the phase velocity. Intuitively the invariant
shape of the soliton solution may then be understood as the consequence
of a delicate balance between nonlinearity and dispersion at higher
orders of both $u$ and $k$ \cite{DrJo}. 

In the present paper we investigate cylindrical solitons of 
KdV-type on the surface of a ferrofluid in the magnetic field of a
current-conducting wire. In order to conserve the radial symmetry of
the problem we neglect gravity. A possible experimental realization
of to this situation is to surround the ferrofluid column with a
non-magnetic fluid of the same density. In this case the hydrodynamics
of this fluid has to be treated as well. 

Ferrofluids are stable suspensions of ferromagnetic nano-particles in
Newtonian liquids and behave superparamagnetically in external
magnetic fields \cite{Ros}. In the standard setup of a horizontal
layer of ferrofluid subject to a homogeneous magnetic field an
additional term proportional to $k^2$ shows up in the dispersion
relation (\ref{eq:disp0}) \cite{CoRo} which inhibits the derivation of
a KdV equation in this geometry. On the other hand, for a ferrofluid 
cylinder in the magnetic field of a current-carrying wire the magnetic
force may replace
gravity and allows for dispersion free surface waves in the long
wavelength limit. This in turn paves the way to derive a KdV equation
for axis-symmetric surface deformations on the ferrofluid cylinder
\cite{foiguel1,foiguel2}. 

Due to surface tension a long fluid cylinder is unstable to surface
modulations resulting eventually in disconnected drops 
(Rayleigh-Plateau instability). Before embarking on the study of
travelling waves on the fluid surface therefore means have to be found
to suppress this instability. Fortunately, this can also be
accomplished with the help of the magnetic field \cite{Ros}.

An accurate experimental investigation of solitons in hydrodynamic 
systems is notoriously difficult due to the ubiquitous presence of
dissipation. Most quantitative studies have been devoted to hump
solitons in shallow channels of water \cite{Zabusky, Hammack} whereas 
recently also the detection of hole solitons on the surface of mercury
have been reported \cite{Falcon}. In our present setup either hole or
hump solitary waves are possible depending on the value of the applied
current. We therefore hope that the present theoretical work may also
stimulate new experimental investigations. 

The paper is organized as follows. In section 2 we collect the basic
equations and boundary conditions. Section 3 is devoted to a linear
stability analysis of a cylinder of ferrofluid in the magnetic field
of a current-carrying wire. Here we demonstrate the possibility to
suppress the Rayleigh-Plateau instability and establish the dispersion
relation (\ref{eq:disp0}) for axis-symmetric surface waves. In section
4 we derive the KdV equation by multiple scale perturbation theory
with details of the calculation relegated to two appendices. Section 5
provides the explicit form of the one- and two-soliton solution and
gives some estimates for possible experimental realizations. Finally,
section 6 contains some conclusions.


\section{Basic equations}

\begin{figure}[h]
  \centering
  \includegraphics[scale=1]{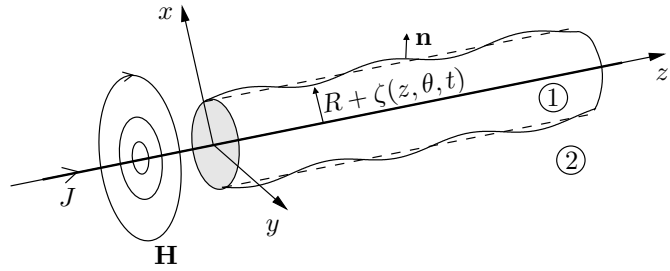}
  \caption{ \label{fig:model} Schematic plot of the system under
    consideration. A current-carrying wire is surrounded by a
    ferrofluid column with magnetic susceptibility $\chi$ 
    and density $\rho$ (region \textcircled{1}) under zero
    gravity. Region \textcircled{2} is a nonmagnetic medium  of
    negligible density treated as vacuum. The dynamics of the
    deflection $\zeta$ of the surface from the perfect  
    cylindrical shape with radius $R$ is the central quantity of
    interest. The vector $\vector n$ denotes the normal on the free
    interface $R+\zeta(z,\theta,t)$.}  
\end{figure}

We consider a cylindrical column of ferrofluid surrounding a straight,
thin, long, current-carrying wire under zero gravity. The ferrofluid
is modelled as an incompressible, inviscid liquid of density $\rho$
and constant magnetic susceptibility $\chi$ surrounded by a vacuum. 
Although we will eventually be interested in the non-linear evolution
of the surface profile of the fluid the assumption of a linear
magnetization law ${\bf M} = \chi {\bf H}$ is quite reasonable for
experimentally relevant parameters as will be discussed in section 5.
We use cylindrical coordinates ($r,\theta,z$), with the $z$-axis
pointing along the wire (see figure~\ref{fig:model}). The magnetic
field is given by  
\begin{equation}
  \label{eq:field}
  {\bf H} = \frac{J}{2\pi r}{\bf e}_\theta\, ,
\end{equation}
where $J$ denotes the current through the wire. Due to the field a
magnetization ${\bf M}$ builds up in the
ferrofluid. The corresponding magnetic force, 
${\bf F}_m = \mu_0 ({\bf  M}\nabla){\bf  H}$ attracts the ferrofluid
radially inward.    
The equilibrium free surface of the ferrofluid is hence cylindrical
with the radius denoted by $R$. Deviations from this shape are 
parametrized by a function $\zeta(z,\theta,t)$ according to
$r=R+\zeta(z,\theta,t)$. 

The velocity field $ {\bf v}(r,\theta,z)$ inside the ferrofluid is
determined by the continuity equation  
\begin{equation}
  \nabla \cdot\vector v = 0 \, ,
\label{eq:cont}
\end{equation}
and by the Euler equation
\begin{equation}
  \label{eq:euler}
  \rho\p_t\vector v + \rho (\vector v\nabla)\vector v = -\nabla P +
  \mu_0 ({\bf M}\nabla) {\bf H}\,.
\end{equation}
Here $P(r,\theta,z)$ denotes the pressure. We will only consider
situations in which the flow of the fluid is irrotational,  
\begin{equation}
  \label{eq:rotv}
  \vector \nabla \times \vector v = 0 \, .
\end{equation} 
It is convenient then to introduce a scalar potential for the
velocity
\begin{equation}
  {\bf v} =\nabla\Phi 
\label{eq:defphi}
\end{equation}
which due to (\ref{eq:cont}) fulfills the Laplace equation
\begin{equation}
  \label{eq:vpot}
  \Delta\Phi =0 
\end{equation}
The Euler equation may now be integrated once to yield the Bernoulli
equation 
\begin{equation}
  \label{eq:bernoulli}
  \rho \partial_t \Phi + \frac{\rho}{2} (\nabla \Phi)^2 
      + P -\frac{\mu_0\chi}{2} H^2 =const. 
\end{equation}

The magnetic field has to obey the magnetostatic Maxwell equations
\cite{Ros} 
\begin{equation}
  \label{eq:maxwell}
  \eqalign{
     \nabla\cdot{\bf H} = 0 \\ 
     \nabla\times{\bf H} = 0 \, ,}
\end{equation}
both inside and outside the ferrofluid. Denoting the respective fields
by ${\bf H}_1$ and ${\bf H}_2$ equations~(\ref{eq:maxwell}) allow the
representations  
\begin{equation}
  {\bf H_1} =-\nabla\Psi_1 \qquad{\rm and}\qquad{\bf H_2} =-\nabla\Psi_2
\label{eq:defpsi}
\end{equation}
with the scalar magnetic potentials $\Psi_1$ and $\Psi_2$ also
fulfilling the Laplace equation:
\begin{equation}
  \label{eq:magpot}
  \Delta\Psi_1 =0 \qquad{\rm and}\qquad \Delta\Psi_2 =0  \, .
\end{equation}

Equations~(\ref{eq:vpot}), (\ref{eq:bernoulli}), and (\ref{eq:magpot})
are to be complemented by boundary conditions.
On the hydrodynamic side we have, assuming no radial extension of
the wire, 
\begin{equation}
  \lim_{r\to 0} \partial_r\Phi = 0 \, .
  \label{eq:draht}
\end{equation}
Moreover at the free surface we need to fulfill the kinematic
condition  
\begin{equation}
  \partial_t\zeta + \partial_z\Phi\partial_z\zeta +\frac{\p_\theta\Phi
  \p_\theta\zeta}{r^2} = \partial_r\Phi
\label{eq:kinematic}
\end{equation}
as well as the pressure equilibrium \cite{Ros}
\begin{equation}
  \label{eq:druck}
  P = P_0 + \sigma K - \frac{\mu_0}{2}M_n^2 \, .
\end{equation}
Here $\sigma$ is the surface tension, $K:=\nabla\cdot \vector{n}$
denotes the curvature of the free surface, and $M_n$ is the
magnetization perpendicular to the surface. The normal vector
$\vector{n}$ on the surface is given by  
\begin{equation}
    \label{eq:normale}
    {\bf n} = \frac{\nabla (r - \zeta(z,\theta,t))}
    {|\nabla (r - \zeta(z,\theta,t))|}\,.
\end{equation}
Note that $\zeta\equiv 0$ yields $K=1/R$ as it should be for the
undisturbed cylinder. 

The boundary conditions for the magnetic field assume the form
\cite{Ros} 
\begin{equation}
  \label{eq:unendlich}
  \eqalign{
     \lim_{r\ra 0} \p_r\Psi_1 = 0\\
     \lim_{r\ra\infty} \p_r\Psi_2 = 0 \, .}
\end{equation}
At the free surface we have 
\begin{equation}
  \label{eq:magnetic-cond}
  \eqalign{
    \vector n \cdot\vector\nabla
          \left(\Psi_2 -(1+\chi)\Psi_1\right) = 0\\
    \Psi_2 - \Psi_1 = 0 \, .}
\end{equation}
Equations~(\ref{eq:magnetic-cond}) describe the feedback of the flow
of the ferrofluid onto the magnetic field.   


It is convenient to introduce dimensionless units. We measure all
lengths in units of the cylinder radius $R$, and use the replacements 
\begin{equation}
     t\ra\sqrt{\frac{R^3\rho}{\sigma}}\,t\, ,\quad
      \Phi\ra\sqrt{\frac{R \sigma}{\rho}}\,\Phi\,,\quad P\ra\frac{\sigma}{R}\, P, \quad 
      \Psi \ra \frac{J}{2\pi} \Psi\, .
\label{dimlos}
\end{equation}
The overall magnetic field strength which can be externally controlled
by changing the current $J$ is then characterized by the dimensionless
magnetic Bond number
\begin{equation}
  \label{eq:defbo}
  Bo:=\frac{\mu_0\chi J^2}{4\pi^2\sigma R}\, .
\end{equation}

Using the Bernoulli equation (\ref{eq:bernoulli}) the pressure
equilibrium (\ref{eq:druck}) at the free surface
$r=1+\zeta(z,\theta,t)$ is given by  
\begin{equation}\nonumber
  \partial_t\Phi + \frac{1}{2}\left(\nabla\Phi\right)^2 +
  \nabla\cdot\vector{n} - \frac{Bo}{2}\left(\chi ({\bf n}\cdot\nabla
  \Psi_1)^2 + (\nabla \Psi_1) ^2\right) = 1-\frac{Bo}{2} \, .
\label{eq:bernoulli1}
\end{equation}
Here the reference pressure $P_0$ in (\ref{eq:druck}) has been chosen
such that $\Phi\equiv 0, \zeta\equiv 0$ is a solution of
(\ref{eq:bernoulli1}).  

\section{Linear stability analysis}

In this section we study the linear stability of the cylindrical
interface given by $\zeta\equiv 0$, $\Phi\equiv 0$ ,
$\Psi_1=\Psi_2=\theta$. To this end we introduce small
perturbations $\zeta(\theta,z,t), \phi(r,\theta,z,t),
\psi_1(r,\theta,z,t)$, and $\psi_2(r,\theta,z,t)$ of the surface
profile, velocity potential and magnetic potentials respectively and
linearize the basic equations and their boundary conditions in these
perturbations. From the translational invariance along the $z$-axis
and equations~(\ref{eq:vpot}) and (\ref{eq:magpot}) together with the
boundary conditions (\ref{eq:draht}) and (\ref{eq:unendlich}) it
follows that these perturbations are of the form
\begin{eqnarray}
  \zeta(\theta,z,t)    &=& C_n \exp(in\theta +ikz + pt)\\
  \phi(r,\theta,z,t)   &=& D_n I_n(kr)\exp(in\theta +ikz + pt)\\
  \psi_1(r,\theta,z,t) &=& A_n I_n(kr)\exp(in\theta +ikz + pt)\\
  \psi_2(r,\theta,z,t) &=& B_n K_n(kr)\exp(in\theta +ikz + pt)\, .
\end{eqnarray}
Here $k$ denotes the wave number in $z$-direction, $n\in \Z$
characterizes the azimuthal modulations, and $p$ is the growth
rate. The $A_n, B_n, C_n$ and $D_n$ are constants (with their
dependence on $k$ and $p$ suppressed) and $I_n(k)$ and $K_n(k)$ denote
modified Bessel functions of order $n$ \cite{AbSt}. 

Using the linearization of (\ref{eq:magnetic-cond}) we may express 
$A_n$ and $B_n$ in terms of $C_n$ according to 
\begin{equation}
  \label{eq:magn-pot}
   \eqalign{
    A_n &= in\chi\frac{K_n(k)}{I_n(k)K_n'(k)-
      \mu_r I_n'(k)K_n(k)}\,C_n \\
    B_n &= in\chi\frac{I_n(k)}{I_n(k)K_n'(k)-
      \mu_r I_n'(k)K_n(k)}\,C_n \, ,}
\end{equation}
where the prime denotes differentiation with respect to the argument. 
In addition the linearized version of (\ref{eq:kinematic}) gives 
\begin{equation}\label{eq:h2}
  D_n = \frac{p}{k I_n'(k)}\,C_n \,.
\end{equation}
Finally, linearizing (\ref{eq:bernoulli1}) we find 
\begin{equation}
  \label{eq:h1}
  \partial_t \phi - \partial^2_\theta \zeta - \partial^2_z \zeta
    + (Bo-1) \zeta +Bo\, \partial_\theta \psi_1=0 
\end{equation}
which when combined with (\ref{eq:magn-pot}) and (\ref{eq:h2}) yields the
dispersion relation
\begin{equation}
  p_n^2(k) = k\frac{I'_n(k)}{I_n(k)}\left(1-n^2-Bo-k^2\right) + \frac{n^2\chi
  Bo}{\frac{I_n(k)K'_n(k)}{I'_n(k)K_n(k)}-(1+\chi)}\, .
\label{eq:disprel}
\end{equation}
The reference state of a cylindrical column becomes unstable if
combinations of $k,n$ and $Bo$ exist for which $p_n$ is positive. 

For $Bo=0$ we find back the well-known Rayleigh-Plateau instability
accomplished by radially symmetric modes with $n=0$. Modes with higher
values of $n$ are not able to destabilize the fluid cylinder. 

\begin{figure}[h]
  \centering
  \includegraphics[scale=0.7]{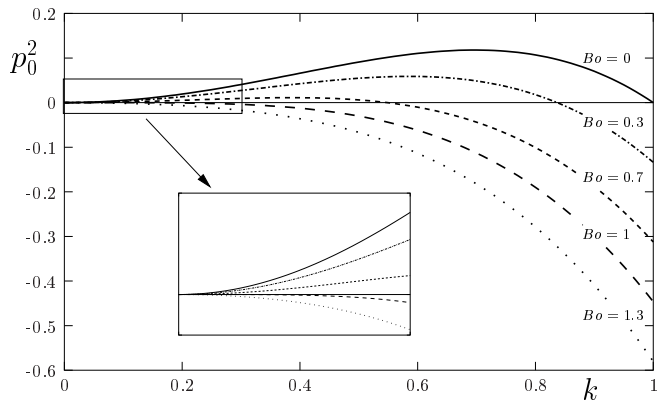}
  \caption{\label{fig:disprel_n=0} Plot of the square of the growth
    rate of axis symmetric distortions as a function of the wave
    number $k$ as given by (\ref{eq:disprel_n=0}) for magnetic Bond
    numbers $Bo=0, 0.3, 0.7, 1$ and $Bo=1.3$ (from top to bottom). 
    The inset shows a magnification of the region around $k=0$. 
    For $Bo>1$ one has $p^2<0$ for all $k$ and no instability occurs.}
\end{figure}

Since one has for all $k$ 
\begin{equation}
  \label{eq:h9}
  \frac{I'_n(k)}{I_n(k)}>0 \qquad{\rm and}\qquad
  \frac{I_n(k)K'_n(k)}{I'_n(k)K_n(k)}<0
\end{equation}
we infer from (\ref{eq:disprel}) that $p^2_n(k)$ is a monotonically
decreasing function of the magnetic Bond number $Bo$. The magnetic
field hence always stabilizes the cylindrical surface. 
Consequently it may change the qualitative behaviour of the system
only due to its influence on the $n=0$ modes. For $n=0$ the dispersion
relation reads 
\begin{equation}
  \label{eq:disprel_n=0}
  p_0^2(k) = k\frac{I_1(k)}{I_0(k)}\left(1-Bo-k^2\right)\, .
\end{equation} 
It is displayed for several values of $Bo$ in 
figure~(\ref{fig:disprel_n=0}).
From (\ref{eq:disprel_n=0}) we see that the Rayleigh-Plateau
instability for a ferrofluid column will be suppressed by 
a sufficiently strong magnetic field fulfilling $Bo>1$ 
\cite{BeBa}. Using typical parameter values as $\chi=1.2$,
$\sigma=0.03\, \mathrm{J/m^2}$ and a fluid radius of 
$R=1$ cm the system remains stable if the current exceeds
the threshold $J_c\simeq 89$ A.
 
It is instructive to investigate the dispersion relation
(\ref{eq:disprel}) in the {\it long wavelength limit} ($k\ll 1$). 
Using the expansion of the modified Bessel functions for small arguments
\cite{AbSt} we get
\begin{eqnarray}
  &\frac{I_0'(k)}{I_0(k)} \sim \frac k 2-\frac{k^3}{16}\\
  &\frac{I_n'(k)}{I_n(k)} \sim \frac n k \quad{\rm if}\quad n>0\\
  &\frac{I_0(k)\,K_0'(k)}{I_0'(k)\,K_0(k)} \sim \frac{2}{k^2 \log k}\\
  &\frac{I_n(k)\,K_n'(k)}{I_n'(k)\,K_n(k)} \sim -1 \quad{\rm if}\quad
  n>0\, ,
\end{eqnarray}
and hence find 
\begin{eqnarray}
  p_0^2(k) &=& \frac{1-Bo}{2}k^2 -\frac{9-Bo}{16}k^4+\order{k^6}\\
  p_n^2(k) &=& n(1-n^2-Bo)-n^2 \frac{\chi Bo}{\chi+2} 
  +\order{k^2}\quad{\rm if}\quad n>0\,.\label{eq:h4}
\end{eqnarray}
Therefore for $n=0$ and $Bo>1$ the system exhibits surface waves
$\zeta(z,t)\sim \exp{i(kz-\omega t)}$ with dispersion relation
\begin{equation}
  \omega(k) = \sqrt{\frac{Bo-1}{2}}\;k
   \left(1-\frac{1}{16}\frac{Bo-9}{Bo-1}\;k^2\right) +\order{k^5}
\label{eq:kdvdisprel}
\end{equation}
The important point for what follows is that these surface waves
become {\it dispersion free}, $\omega=c_0 k$, in the long
wavelength limit $k\to 0$. The phase velocity is given by 
\begin{equation}
  \label{eq:defc0}
  c_0=\sqrt{\frac{Bo-1}{2}}\, .
\end{equation}
The situation is hence analogous to the shallow water
equations which form the starting point for the derivation of the
Korteweg-de~Vries equation in a rectangular duct \cite{whitham}. Note that
no such waves are possible for $n>1$, cf. (\ref{eq:h4}).


\section{Korteweg-de~Vries equation}

In the previous section we have seen that the system under
consideration admits to linear order in the
surface deflection $\zeta$ cylindrical, axis symmetric surface waves
with no dispersion in the long wavelength limit $k\to 0$. From the
experience with plane surface waves on shallow water \cite{whitham} it is
hence tempting to investigate whether at higher orders in $k$ and 
$\zeta$ nonlinear waves may be obtained for which the effects of
nonlinearity and dispersion exactly balance each other. This could
then give rise to axis symmetric soliton solutions in the present
cylindrical geometry.   

In this section we show that it is indeed possible to derive a KdV
equation for the surface deflection $\zeta(z,t)$ 
\cite{foiguel1} by using a multiple scale perturbation theory similar to
the case of rectangular geometry. To this end we first observe that
for an axis symmetric free surface the magnetic field problem
decouples from the hydrodynamics and we have the exact result
\begin{equation}
  \label{eq:psi1}
  \Psi_1=\theta \, .
\end{equation}
This in turn implies $\nabla \Psi_1=(0,1/r,0)$ and therefore
\mbox{${\bf n}\cdot\nabla \Psi_1=0$}. Using moreover the explicit expression
for ${\bf n}$ in terms of $\zeta(z,t)$ resulting from
(\ref{eq:normale}) we get from (\ref{eq:bernoulli1}) 
\begin{equation}
  \label{eq:bernoulli3}
  \fl
   \p_t\Phi + \frac{1}{2}\left[(\p_z\Phi)^2 + (\p_r\Phi)^2\right] +
    \frac{\frac{1+(\p_z\zeta)^2}
      {1+\zeta} -
      \p_z^2\zeta}{\left[1+(\p_z\zeta)^2\right]^{\frac{3}{2}}}
    - \frac{1}{2}\frac{Bo}
         {(1+\zeta)^2} = 1-\frac{Bo}{2}\, .
\end{equation}
The kinematic condition (\ref{eq:kinematic}) simplifies to 
\begin{equation}
  \partial_t\zeta + \partial_z\Phi\partial_z\zeta = \partial_r\Phi
\label{eq:kinematic1}
\end{equation}

The KdV equation appears in the limit of small nonlinearity, 
$\zeta\ll 1$, and small dispersion, $k\ll 1$ with the  proper balance
between these two ingredients occurring for $\zeta=\order{k^2}$. To
make this combined limit explicit we introduce a small parameter, 
$\eps$, and use the rescalings
\begin{equation}
  \label{eq:skalen}
  z\rightarrow\frac{z}{\sqrt{\eps}}\,,\quad r\rightarrow r\,,\quad
  \zeta\rightarrow\epsilon\zeta\,,\quad t\rightarrow\frac{t}{c_0\sqrt{\eps}}\,,\quad
  \Phi\rightarrow\sqrt{\eps}\, c_0\,\Phi \, ,
\end{equation}
where $c_0$ is defined by (\ref{eq:defc0}).  To
derive the KdV equation we will need the two basic equations
(\ref{eq:bernoulli3}) and (\ref{eq:kinematic1}) up to order
$\eps^2$. Plugging (\ref{eq:skalen}) into these equations we find to
the required order 
\begin{equation}
  \label{eq:bernoulli4} 
    \p_t\Phi + \frac{1}{2}\left[\eps\,(\p_z\Phi)^2 +
      (\p_r\Phi)^2\right] + 2\zeta - \frac{\eps}{2c_0^2}(3Bo -2)\, \zeta^2 - 
       \frac{\eps}{c_0^2}\p_z^2\zeta =0
\end{equation}
and 
\begin{equation}
  \label{eq:kinematic1a}
  \eps\, \p_t \zeta + \eps^2 \,\p_z\Phi\,\p_z\zeta=\p_r\Phi \, .
\end{equation}
To get a suitable expansion for the velocity potential $\Phi$ we note
that from the Laplace equation (\ref{eq:vpot}) and the boundary
condition (\ref{eq:draht}) one may derive the 
following representation for $\Phi(r,z,t)$ (see Appendix A) 
\begin{equation}
  \label{eq:summephi}
    \Phi(r,z,t) = \sum_{m=0}^\infty r^{2m} \,\eps^m \,
  \frac{(-1)^m}{(2^m m!)^2}\;\partial_z^{2m}\Phi_0(z,t)
\end{equation}
with the so far undetermined function $\Phi_0(z,t)$. 

Using this expansion for $\Phi$ in (\ref{eq:bernoulli4}) and
(\ref{eq:kinematic1a}) and observing that both equations hold at the
interface, i.e. for $r=1+\eps\,\zeta$, we get to the desired order in
$\eps$ 
\begin{equation} 
 \label{eq:bernoulli5}
 \p_t\Phi_0 +2\zeta=
    \frac{\eps}{4}\p_t\p_z^2 \Phi_0
       -\frac{\eps}{2}(\p_z\Phi_0)^2 + 
       \frac{\eps(3Bo -2)}{2c_0^2}\zeta^2 + \frac{\eps}{c_0^2}
       \p_z^2\zeta 
\end{equation}
and
\begin{equation}
  \label{eq:kinematic2}
  \p_t\zeta + \frac1 2 \p_z^2\Phi_0=
 -\eps\p_z\Phi_0\,\p_z\zeta
    -\frac{\eps}{2}\zeta\,\p_z^2\Phi_0 + 
      \frac{\eps}{16}\p_z^4\Phi_0
\end{equation}
It is convenient to differentiate (\ref{eq:bernoulli5}) with respect
to $z$ and to introduce the $z$-component of the velocity of the
ferrofluid $u=\p_z\Phi$. We then find the final set of equations to
determine $\zeta$ and $u$ 
\begin{equation} 
 \label{eq:bernoulli6}
 \p_t u +2\p_z\zeta=\eps\left(\frac{1}{4}\p_t\p_z^2 u -u\p_z u + 
  \frac{3Bo-2}{c_0^2}\zeta \p_z \zeta + \frac{1}{c_0^2} \p_z^3\zeta \right)
\end{equation}
and
\begin{equation}
  \label{eq:kinematic4}
  \p_t\zeta + \frac1 2 \p_z u= \eps \left( -u\p_z\zeta
    -\frac{1}{2}\zeta\,\p_z u + \frac{1}{16}\p_z^3 u \right) \, .
\end{equation}
We now solve these equations perturbatively using the ans\"atze 
\begin{equation}
  \eqalign{
    \zeta(z,t,\tau) = \zeta_0(z,t,\tau) +\epsilon\zeta_1(z,t,\tau) 
      + \order{\epsilon^2}\\
    u(z,t,\tau) = u_0(z,t,\tau) + \epsilon u_1(z,t,\tau) 
   + \order{\epsilon^2}\, ,}
  \label{potenz}
\end{equation}
where we have introduced a second, slow time variable $\tau:=\eps t$. 
Plugging these expansions into (\ref{eq:bernoulli6}) and
(\ref{eq:kinematic4}) we find to zeroth order in $\eps$
\begin{equation}\label{ordnungnull}
    L\left(\begin{array}{c}
      u_0\\[1ex]
      \zeta_0\end{array}\right) = 
     \left(\begin{array}{c} 0\\[1ex] 0 \end{array}\right)
\end{equation}
where the linear operator $L$ is given by
\begin{equation}
  \label{eq:defL}
  L=\left(\begin{array}{cc}
        \partial_t & 2\partial_z\\[1ex]
        \frac{1}{2}\partial_z & \partial_t\end{array}\right)\, .
\end{equation}
The solution are dispersion free travelling waves of d'Alembert
form 
\begin{equation}
  \label{zeta+u}
  \eqalign{
    u_0(z,t,\tau) &= 2f(z-t,\tau)\\
    \zeta_0(z,t,\tau) &= f(z-t,\tau)\, ,}  
\end{equation}
with a so far unspecified function $f(x,\tau)$ where we have
restricted ourselves to waves travelling to the right. 

To order $\eps$ we find 
\begin{equation}
  \fl
    \label{eq:order1}
    L\left(\begin{array}{c}
      u_1\\[1ex]
      \zeta_1\end{array}\right) = \left(\begin{array}{c}
          -\partial_\tau u_0 + \frac{1}{4}\partial_t\partial_z^2u_0
          - u_0\partial_z u_0 + \frac{3Bo-2}{c_0^2}\zeta_0\partial_z\zeta_0
          + \frac{1}{c_0^2}\partial_z^3\zeta_0\\[1ex]          
          -\partial_\tau\zeta_0 - u_0\partial_z\zeta_0 -
          \frac{1}{2}\zeta_0\partial_z u_0 + \frac{1}{16}\partial_z^3
          u_0
        \end{array}\right)\, .
\end{equation}
This inhomogeneous equation involves again the linear operator $L$
which is singular, cf. (\ref{ordnungnull}). Hence the
inhomogeneity of this equation must be orthogonal to the zero eigenspace of
the adjoint operator $L^+$. The determination of $L^+$ and the
projection of the r.h.s. of (\ref{eq:order1}) onto the eigenfunction
of $L^+$ with eigenvalue zero is done in appendix B. The solvability
condition for (\ref{eq:order1}) finally acquires the form 
\begin{equation}
  \label{eq:kdv}
  \p_\tau f + \frac{2Bo-3}{4c_0^2}f\p_zf + \frac{Bo-9}{32c_0^2}\p_z^3 f = 0
\end{equation}
Using (\ref{zeta+u}), denoting $\zeta_0$ simply by $\zeta$  and
reversing the scalings (\ref{eq:skalen}) then 
yields the following KdV equation for the surface deflection
$\zeta(z,t)$
\begin{equation}
  \label{eq:kdv-galilei}
  \p_t \zeta + c_0 \p_z\zeta +\frac{2Bo-3}{4c_0}\,\zeta \p_z\zeta + 
           \frac{Bo-9}{32c_0}\,\p_z^3 \zeta  = 0
\end{equation}
When discussing the implications of this equations one has to keep in
mind that it is valid for small $\zeta$ only. 


\section{Results}

Equation~(\ref{eq:kdv-galilei}) is of the form 
\begin{equation}\label{eq:kdvgen}
  \p_t \zeta + c_0 \p_z\zeta +c_1\,\zeta \p_z\zeta + 
           c_2\,\p_z^3 \zeta  = 0 \, ,
\end{equation}
with the coefficients 
\begin{equation}
  c_0=\sqrt{\frac{Bo-1}{2}}\,,\quad
  c_1=\frac{2Bo-3}{4c_0}\,,\quad{\rm and}\quad 
  c_2=\frac{Bo-9}{32c_0}
\end{equation}
all depending on the magnetic field strength $Bo$. From section III we
know that we must have $Bo>1$ since otherwise the fluid cylinder is
susceptible to the Rayleigh-Plateau instability. Hence both $c_1$ and
$c_2$ may change sign for allowed values of $Bo$. 

\begin{figure}[h]
  \centering
  \includegraphics[scale=0.35]{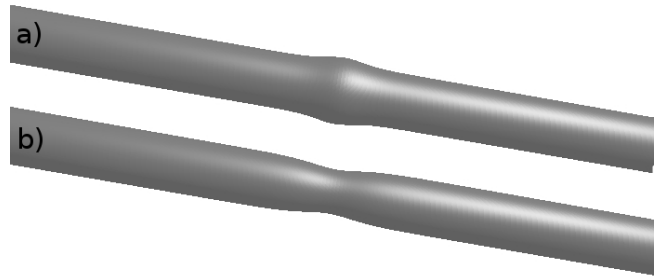}
  \caption{\label{fig:soliton} Schematic plot of a cylindrical
    hump-soliton (a) and hole-soliton (b). The parameter values are
    specified in the main text.}
\end{figure}

The one-soliton solution of (\ref{eq:kdvgen}) is of the form
(cf.~(\ref{eq:onesol})) 
\begin{equation}
  \label{eq:kdvsol}
  \zeta^{(1)}(z,t) = \frac{3c}{c_1} 
   \mbox{sech}^2\left(\sqrt{\frac{c}{4c_2}}(z-(c+c_0)t)\right)\, .
\end{equation}
where $c\ll 1$ is a free constant having the same sign as $c_2$. For
$Bo<9$ we have hence $c<0$ and the soliton has a slightly smaller 
velocity than the linear waves. If $Bo<3/2$ also $c_1<0$ and therefore
the amplitude of the soliton is positive, i.e. we have a hump soliton
as shown in figure~\ref{fig:soliton}a. For $3/2<Bo<9$ on the other hand 
$c_1>0$ and consequently (\ref{eq:kdvsol}) describes a depression or
hole soliton as depicted in figure~\ref{fig:soliton}b. Finally, for
$Bo>9$ we have $c_2>0$, hence $c>0$, and also $c_1>0$. The soliton
amplitude is therefore positive again and its velocity is now slightly
larger than that of the corresponding linear waves. 

To get some impression of the accessibility of the solution in
experiments the results for the following parameter sets may be
helpful. For a ferrofluid with $\chi=1.2$, $\rho=1.12\,{\rm g/cm^{3}}$
and $\sigma=0.03\, \mathrm{J/m^2}$
forming a cylinder of radius $R=1\,\mathrm{cm}$  a current
$I=100\,\mathrm{A}$ corresponds to $Bo\simeq 1.27$. A soliton with
amplitude $A=2\,\mathrm{mm}$ has then a velocity of
$U=1.8\,\mathrm{cm/s}$ and the width of the hump is about
$L=20\,\mathrm{cm}$. This soliton will 
hence be difficult to observe in an experiment. For a current of
$I=294\,\mathrm{A}$ corresponding to $Bo\simeq 11$ the extension
reduces for the same amplitude to $L=1.6\,\mathrm{cm}$ with the velocity
increasing to $U=12.3\,\mathrm{cm/s}$. A hole-soliton with amplitude
$A=-2\,\mathrm{mm}$, velocity $U=8.4\,\mathrm{cm/s}$, and width
$L=2.1\,\mathrm{cm}$ can be realized with a current of $I=235\,\mathrm{A}$
corresponding to $Bo\simeq 7$. The latter two solitons are shown
schematically in figure~\ref{fig:soliton}. Both should be easily
observable experimentally. Note that for such values of the current the magnitude of error assuming
$\chi =const.$ of the magnetization is about 5~\%. 

A two-soliton solution may be derived using, e.g., Hirotha's method
\cite{hirota}. Depending on the value of the magnetic bond number one
may combine either two hump or two hole solitons. The case of two hump
solitons is described by the solution 
\begin{equation}
\fl
  \label{eq:2kdvsoliton}
    \zeta^{(2)}(z,t)= 8\frac{\gamma_1^2\xi_1 + \gamma_2^2\xi_2 +
      \left(\gamma_1-\gamma_2\right)^2\xi_1\xi_2  
      + \left(\frac{\gamma_1-\gamma_2}{\gamma_1+\gamma_2}\right)^2\!
      \left(\gamma_1^2\xi_1\xi_2^2 +
      \gamma_2^2\xi_1^2\xi_2\right)}{\left(1+\xi_1 + \xi_2 +
      \left(\frac{\gamma_1-\gamma_2}
           {\gamma_1+\gamma_2}\right)^2\!
           \xi_1\xi_2\right)^2} 
\end{equation} 
where
\begin{equation}
    \gamma_i^2 = \frac{3 c_i}{c_1} \quad{\rm and}\quad
    \xi_i =  \exp\left(\sqrt\frac{c_i}{c_2}(z-z_{0i}-(c_i+c_0)t)\right)
\end{equation}
for $i=1,2$. 

\begin{figure}[h]
  \centering
  \includegraphics[scale=0.6]{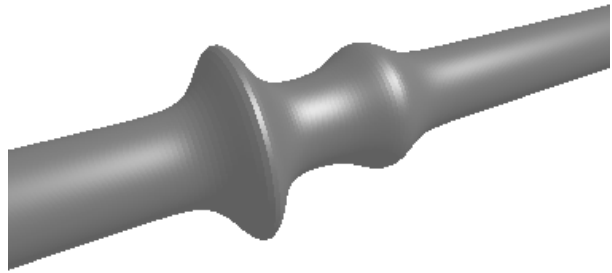}
  \caption{\label{fig:2kdvsoliton} Schematic snapshot of the
    two-soliton solution described by eq.~(\ref{eq:2kdvsoliton}) for the
    parameter values $R=1.8\,{\rm cm}$, $Bo=11$, $A_1=8\,{\rm mm}$,
    $A_2=4\,{\rm mm}$ at $t=-0.5$.}
\end{figure}

\begin{figure}[h]
  \centering
  \includegraphics[scale=0.9]{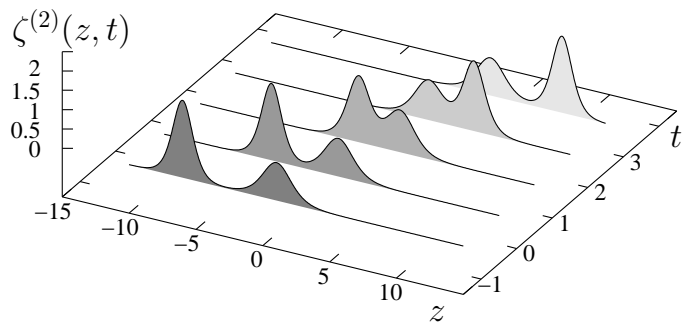}
  \caption{\label{fig:2kdvsoliton1} Evolution of the two-soliton
    solution eq.(\ref{eq:2kdvsoliton}) at with time 
    $t=\{-1.5,-0.5,0.5,1.5,2.5\}$. The parameters are the same as in
    figure~\ref{fig:2kdvsoliton}}
\end{figure}

A snapshot of the solution is displayed in figure~\ref{fig:2kdvsoliton}, its
time evolution is characterized by figure~\ref{fig:2kdvsoliton1}. The
main feature is the passing of the slower soliton by the faster
one. After the interaction process the two solitons reemerge
undisturbed which is the defining property of a soliton solution. An
animated version of the two-soliton solution is shown in a movie for
parameters values as in figure~\ref{fig:2kdvsoliton}.



\section{Conclusion}

In the present paper we have investigated nonlinear waves on the
cylindrical surface of a ferrofluid surrounding a current-carrying
wire under zero gravity. We have shown that for a sufficiently large
current a Korteweg-de~Vries equation for axis-symmetric surface
distortion can be derived. Accordingly the system under consideration
is well suited to experimentally investigate cylindrical solitons of
KdV type.  

In order to observe these solitons first of all the ubiquitous
Rayleigh-Plateau instability has to be suppressed. This can be 
accomplished by the magnetic field if the current exceeds a critical
value which for experimentally relevant parameters is of about 100 A. 

We have shown that non axis-symmetric perturbations of the
surface always disperse whereas axis-symmetric ones propagate
almost dispersion-free if the wavelength is very large. Using the fact
that for axis-symmetric surface deflections the magnetic field problem
decouples completely from the hydrodynamic part a Korteweg-de Vries
equation can be derived. The 
parameters in this equation depend on the magnetic field strength
which gives rise to qualitatively different soliton solutions like
hump and hole solutions for different values of the current in the
wire. The one- and two-soliton solutions were discussed in detail and
conditions for their experimental realization were given. 

It should be noted that several approximations were used in our
theoretical analysis. First of all the derivation of the KdV-equation
is perturbative and therefore approximate as is typical for the
theoretical discussion of solitons in hydrodynamic
systems. Furthermore we have neglected the hydrodynamic influence of
a non-magnetic fluid surrounding the magnetic column which is necessary to
ensure zero gravity in experiments. Finally, viscosity was
neglected throughout the analysis since the Korteweg-de Vries
equation results from the Euler equation describing inviscid
fluids. In experiments one  will hence always see a damping of the
soliton solutions with time due to dissipation by viscous shear
flow \cite{Zabusky,Hammack,Falcon}. It is an attractive idea to
counter-balance these viscous losses
by appropriately chosen time dependent magnetic fields, however we
were not able to find a suitable geometry for this idea to become
operative. In any case a theoretical analysis aiming at this goal has
to go beyond the quasi-static version of ferro-hydrodynamics employed
in the present analysis and has to include magneto-dissipative
couplings, see e.g. \cite{ShMo}. 

We finally note that our system is an experimentally accessible
realization of the introductory example for a soliton given in
chapter 1.4 of the book by Lamb  
\cite{lamb}. There an incompressible fluid inside a cylinder made of
independent elastic rings is considered. The rings are supposed to
deform axis-symmetrically in reaction to the fluid pressure. However,
although confining the liquid tightly they must be uncoupled in order
not to sustain elastic waves by themselves. Gravity is neglected
altogether. Using the conservation of mass and momentum of the fluid
and linear elasticity for the rings it is then possible to derive a
KdV equation for axis-symmetric deformations of the rings. As we have
shown in the present paper the somewhat unrealistic properties of 
the elastic rings can be mimicked by a cylindrical magnetic field if
the fluid to be confined is a ferrofluid.\\

\ack
We are indebted to Konstantin Morozov for
fruitful discussions and for pointing out references
\cite{foiguel1,foiguel2} to us. We would also like to thank
S. Gro{\ss}mann for a clarifying discussion. 


\appendix

\section{Expansion of $\Phi(r,z,t)$}
After the rescalings (\ref{eq:skalen}) the Laplace equation for the
velocity potential $\Phi$ takes the form
\begin{equation}
  \label{eq:laplace-zyl}
  \frac{1}{r}\partial_r(r\partial_r \Phi(r,z,t)) + 
  \eps \, \partial_z^2 \Phi(r,z,t) = 0
\end{equation}
Representing $\Phi$ as a power series in $r$ 
\begin{equation}
  \label{eq:ansatz_phi}
  \Phi(r,z,t) = \sum_m r^m\Phi_m(z,t)
\end{equation}
we find 
\begin{equation}
  \sum_m r^m\left[(m+2)^2 \Phi_{m+2} + \eps\, \partial^2_z
    \Phi_m\right] = 0
  \label{eq:rekursion1}
\end{equation}
leading to the recursion relation
\begin{equation}
  \label{eq:bedingung}
  \Phi_{m+2} = -\eps \frac{\partial_z^2\Phi_m}{(m+2)^2}\quad.
\end{equation}
Because of the boundary condition (\ref{eq:draht}) we have
\begin{equation}
  \sum_m m\, r^{m-1}\Phi_m = 0
\end{equation} 
implying $\Phi_1=0$. From (\ref{eq:bedingung}) we hence find 
$\Phi_{2m+1} = 0$ for all $m$. The velocity potential may therefore be
expressed entirely in terms of $\Phi_0$ and its derivatives
\begin{eqnarray}
    \Phi(r,z,t) &=& \sum_{m=0}^\infty r^{2m} \,\eps^m \,
  \frac{(-1)^m}{(2^m m!)^2}\;\partial_z^{2m}\Phi_0(z,t)\\
   &=&\Phi_0(z,t)-\frac{\eps r^2}{4} \p_z^2 \Phi_0(z,t)
     +\frac{\eps^2 r^4}{64} \p_z^4 \Phi_0(z,t) +\order{\eps^3} \, .
\end{eqnarray}
which coincides with (\ref{eq:summephi}).

\section{The solvability condition}
Under the usual scalar product 
\begin{equation}
  \label{eq:skalarprodukt1}
  \langle\bar{\Psi}|\Psi\rangle =
  \lim_{Z,T\to\infty}\frac{1}{4ZT}
     \int^Z_{-Z}{\rm  d}z\int^T_{-T}{\rm d}t\; \bar\Psi^*\cdot\Psi  
\end{equation}
with $\Psi=(u,\zeta)$ we find for $L^+$  
\begin{eqnarray}
  L^+ = \left(\begin{array}{cc}
      -\p_t & -\frac{1}{2}\p_z \\[1ex]
      -2\p_z & -\p_t
    \end{array}\right)
\end{eqnarray}
The complete eigenmode to zero eigenvalue of $L^+$ is hence given by
\begin{equation}
  \label{ubar}
  \eqalign{
    \bar{u}_0(z,t) = \bar{f}(z-t)-\bar{g}(z+t)\\
    \bar{\zeta}_0(z,t) = 2(\bar{f}(z-t)+\bar{g}(z+t))\, ,}  
\end{equation}
where $\bar{f}$ and $\bar{g}$ are arbitrary functions of a single
argument. Setting the projection of the r.h.s. of (\ref{eq:order1}) on
this mode equal to zero we find 
\begin{eqnarray}
  \label{eq:hsolve}
\fl
  0 & = \lim_{Z,T\to\infty}\frac{1}{4ZT}
  \int^Z_{-Z}\!{\rm d}z\int^T_{-T}\!{\rm d}t
  \Bigg[\Bigg(-2\p_\tau f +\frac{Bo}{c_0^2}f\p_z f
    -\frac{Bo-5}{4c_0^2}\p_z^3 f\Bigg)
    \Bigg(\bar{f}-\bar{g}\Bigg)\nonumber\\[1ex]
    &\rule{5cm}{0mm}+ 2\Bigg(-\p_\tau f-3f\p_z f +\frac1 8 \p_z^3 f\Bigg)
    \Bigg(\bar{f}+\bar{g}\Bigg)\Bigg]\nonumber\\[2ex] 
  & = \lim_{Z,T\to\infty}\frac{1}{4ZT}
  \int^Z_{-Z}\!{\rm d}z\int^T_{-T}\!{\rm d}t
  \Bigg[\Bigg(-4\p_\tau f
    -\frac{2Bo-3}{c_0^2}f\p_z f-\frac{Bo-9}{8c_0^2}\p_z^3 f\Bigg)\bar{f}\nonumber\\
    &\rule{5cm}{0mm}+ \Bigg(\frac{4Bo-3}{c_0^2}f\p_z f + \frac{3Bo-11}{8c_0^2} \p_z^3 f \Bigg)
    \bar{g}\Bigg]\; .
\end{eqnarray}
The part of the integrand involving $\bar{g}$ may be written in the
form $\bar{g}(z+t)\p_z F(z-t,\tau)$. Substituting
\mbox{$\xi=z-t,\eta=z+t$} one realizes that these terms do not
contribute for $Z,T\to\infty$. Since moreover $\bar{f}$ is an
arbitrary function of its argument (\ref{eq:hsolve}) implies
(\ref{eq:kdv}). 


\end{document}